\documentclass[twocolumn,showpacs,amsmath,amssymb,aps,prl]{revtex4}
\begin{document}

\title{QUANTUM NONLOCALITY AND QUANTUM DYNAMICS}
\author{S. Gheorghiu-Svirschevski\footnotemark[1]\footnotetext{e-mail: hnmg@soa.com}}
\address{1087 Beacon St., Suite 301, Newton, Massachusetts 02459}
\date{\today}

%\maketitle

\begin{abstract}
We argue that usual quantum statics and the dynamical equivalence of mixed quantum states to {\it probabilistic mixtures }suffice to guarantee a linear evolution law and compliance with relativity. Alternatively, there are well-behaved nonlinear dynamics that treat mixed states as {\it elementary mixtures }and evolve {\it every }pure state linearly and unitarily. For these situations the linear evolution of entangled pure states provides an unequivocal signature of linear quantum dynamics.
\end{abstract}
\pacs{03.65.Ta, 03.65.Ud }

\maketitle

\subsection{Introduction}
\label{Sec1}
It was recently claimed in ref.\cite {Gisin} that the usual "quantum statics" supplemented by the no-signaling condition implies a completely positive, linear quantum dynamical law. The proof offered therein relies on the central notion that a mixed quantum state described by a density matrix $\hat\rho$ is always equivalent to, or can be prepared as, a {\it classical probabilistic mixture} of pure states $\left \{ {p_i, |\psi_i\rangle} \right \}$, where $p_i$ denotes the probability of occurrence of the pure state $|\psi_i\rangle$. This assumption was criticized in ref.\cite{Bona} as leading directly to linearity, regardless of the no-signaling condition. We wish to give an expounded version of this criticism, and reexamine the role played by quantum nonlocality in enforcing dynamical linearity. We also review the issue of seemingly viable nonlinear alternatives, particularly the interesting class of nonlinear dynamics that retains the quantum theory of pure states in its linear form. For these cases we point out that the no-signaling condition leads to a simple criterion of dynamical (non)linearity: {\it Quantum dynamics is (non)linear if and only if the entangled pure states of noninteracting systems evolve (non)linearly }.

\subsection{No-signaling and the linearity of quantum dynamics}
\label{Sec2}
Let us begin by recalling the argument of ref.\cite{Gisin}. Consider an arbitrary quantum system, prepared in a mixed quantum state represented by the {\it probabilistic mixture} $\left \{ {p_i, |\psi_i\rangle} \right \}$. If the density matrix corresponding to this mixture reads 
\begin{equation}
\label{eq1}
\hat \rho = \sum\limits_i {p_i |\psi_i \rangle \langle \psi_i | }\;,
\end{equation}
\noindent a dynamical map $\hat\rho \to {\bf g}(\hat\rho)$, not necessarily linear, decomposes as a superposition of individual evolutions of its pure state components $|\psi_i\rangle$. In other words,
\begin{equation}
\label{eq2}
{\bf g}(\hat\rho) \equiv g\left({ \sum\limits_i {p_i |\psi_i \rangle \langle \psi_i | } }\right) = \sum\limits_i {p_i g \left({ |\psi_i\rangle \langle \psi_i | }\right) }\;,
\end{equation} 
\noindent where ${\bf g} \left({ |\psi_i\rangle \langle \psi_i | }\right)$ need not be a pure state. Since this identity must apply to every probabilistic mixture described by the density matrix $\hat \rho$, it is concluded that the dynamical map ${\bf g}$ must be linear. 

The no-signaling condition enters the proof via an argument regarding the reduced state of a quantum system in an entangled pure state with a space-like separated counterpart. Since such a state cannot be prepared locally, one may question whether it represents an essentially indecomposable quantum state, i.e., an {\it elementary mixture} \cite{Bona-1,Gyftopoulos} [although this terminology is not employed in ref.\cite{Gisin}]. On the other hand, the assumption of a potentially nonlinear dynamics necessarily implies the possibility that {\it elementary }and {\it probabilistic mixtures }may evolve differently. Suppose the latter claim is true, and [{\it elementary}] mixtures generated as reduced local states corresponding to entangled pure states may evolve differently than {\it probabilistic mixtures} with identical density matrices. In ref.\cite{Gisin} it is shown that such a reduced state can be converted into any genuine {\it probabilistic mixture} with an identical density matrix by a remote maximal measurement on the entangled counterpart. If the no-signaling condition must apply, then the reduced state must necessarily evolve in the same linear manner as a {\it probabilistic mixture}, or the remote measurement could be detected locally. An additional argument concerning the {\it complete positivity} of the evolution concludes the proof. 

Nevertheless, the above recourse to the no-signaling condition is not necessary once one accepts the interpretation of Eq.(\ref{eq2}) that {\it probabilistic mixtures }must evolve linearly. It is sufficient to note instead that a similar statement must apply to {\it probabilistic mixtures }of pure entangled states [entangled mixtures] of two noninteracting systems, say A and B. In a limiting case, the same can be said about all uncorrelated states of the form $\hat\rho_{A+B} = \hat\rho_A \otimes \hat\rho_B$. If we require now that the uncorrelated states of noninteracting systems must evolve into uncorrelated states, it follows necessarily that the overall dynamical map $g_{A+B}$ [which is linear] must be a direct product of local dynamical maps, ${\bf g}_{A+B} = {\bf g}_A \otimes {\bf g}_B$. Moreover, each of the local maps ${\bf g}_A$ and ${\bf g}_B$ must be linear, by the same line of reasoning derived from Eq.(\ref{eq2}). Since the linear map ${\bf g}_A \otimes {\bf g}_B$ must apply to {\it entangled }pure states as well, one must conclude that the corresponding reduced states also evolve linearly, whether they are {\it probabilistic mixtures} or not.   

At this point one is led to reconsider the motivation behind the claim of ref.\cite{Gisin}. The original observation was that any attempt "to modify quantum physics, e.g., by introducing nonlinear evolution laws for pure states, ...easily leads to the possibility of superluminal communication" [\cite{Gisin}, pg.1]. From this it was reasoned that, perhaps, the converse may be true, and linearity is secured by the relativistic impossibility of superluminal signaling [the no-signaling condition]. As pointed out in ref.\cite{Bona}, and emphasized in the preceding paragraph, this is not the case. Instead, the crucial conjecture leading to linearity is seen to be the postulated identification of mixed quantum states as {\it probabilistic mixtures }in the dynamical sense implicit in Eq.(\ref{eq2}), or alternatively, the dynamical equivalence of {\it probabilistic }and {\it elementary }quantum mixtures. This equivalence proves sufficient to support the no-signaling condition, but the latter does not seem to be a prerequisite for linearity.

The term "dynamical equivalence" is used here to emphasize a distinction from a mere "static equivalence". We reserve the latter term to mean that any maximal measurement on a given mixed state can be reproduced identically by a similar measurement on a unique "probabilistic mixture" of orthogonal pure states, as well as on any member of an [infinite] family of statistically equivalent "probabilistic mixtures" of nonorthogonal states. This is a purely statistical statement, which in itself does not imply a "wavefunction collapse" during measurement, and thus need not involve the "projection postulate". In contrast, a "dynamical equivalence" states that a "statically equivalent" {\it probabilistic mixture} at time $t$ is equivalent to the evolved by pure state propagation of the "statically equivalent" {\it probabilistic mixture} from an earlier time $t_0<t$. Note that a "static equivalence" does not require {\it a priori} a "dynamical equivalence". The reason is simply that the physical behavior of a system is not defined only by its statistics, but also by interactions and, perhaps, self-interactions. In particular, the "static equivalence" of mixed states to {\it probabilistic mixtures} cannot assert the absence or presence of state-dependent self-interactions, and cannot rule out dynamical nonlinearity on its own. On the other hand, the "dynamical equivalence" expressed by Eq.(\ref{eq2}) testifies precisely to the absence of any such self-interactions. 

If we accept this distinction between static and dynamic equivalence, we may resolve that a corrected version of the theorem proposed in ref.\cite{Gisin} can read: \\

{\it Usual quantum statics and the dynamical equivalence of mixed quantum states to probabilistic mixtures implies a linear, completely positive quantum evolution law, which necessarily complies with the no-signaling condition.}  \\ 

This leaves us with the question why exactly nonlinear quantum evolution laws appear incompatible with the no-signaling condition \cite{nlin-ftl-1,nlin-ftl-2,nlin-ftl-3,Polchinski,Mielnik}. The answer remains essentially with the remote preparation procedure discussed in ref.\cite{Gisin}. It is not necessarily nonlinearity itself that is incompatible with no-signaling, but the induced distinguishability of {\it probabilistic }and {\it elementary }mixtures set against intrinsic quantum nonlocality. That is, if {\it elementary mixtures }can be converted by action-at-a-distance into {\it probabilistic mixtures}, then any local means to distinguish between the two types of mixtures, such as a nonlinear dynamics, or the nonlinear observables discussed in refs.\cite{Polchinski,Mielnik}, necessarily opens the possibility of superluminal communication. 

\subsection{Nonlocality and nonlinear quantum dynamics}
\label{Sec3}

One may object that the concept of remote preparation leading to the above conclusion involves an implicit acceptance of the "projection postulate" \cite{Bona}, contrary to the claim of ref.\cite{Gisin} that a strict statistical interpretation of nonlocal correlations should suffice. The reason lies in the intrinsic interconnection between the "probabilistic mixture" interpretation of mixed quantum states, in particular of the output states of quantum measurements, and the "projection postulate". For suppose again that two noninteracting systems $A$ and $B$, at rest relative to each other, are initially in an entangled pure state $|\Psi_{AB}\rangle$. Their respective local states are necessarily "elementary mixtures" that cannot be prepared locally. Let a maximal preparatory measurement be effected and recorded on system $B$ at time $t_0$ as observed in the common rest frame, and let system $A$ be subject to an observation, also immediately recorded, at an arbitrarily close time $t_0+\delta t$. Then if the state of system $A$ is to be identified at $t_0+\delta t$ as a genuine "probabilistic mixture", amenable at least in principle to local preparation, it follows that the remote preparation at $t_0$ must transform the joint state of $A$ and $B$ into a separable mixture, realizable as a statistical superposition of products of local states [otherwise the claim that the state of $A$ is no longer an "elementary mixture" cannot be upheld]. In other words, the total state must undergo a {\it disentanglement process}. We note in passing that such a disentanglement transformation will leave in general the measured system $B$ entangled with a measuring device, in which case the total state of $(A + B + {\rm measuring \; device})$ should become a separable state of $(B + {\rm measuring \; device})$ and $A$.  Now, if at $t_0+\delta t$ this total state of {\it separable form} is indeed a "probabilistic mixture", then by the very definition of the latter every pair of $A$ and $B$ [${\rm + measuring \; device}$] in an ensemble representative of this mixture must actually exist in one of the contributing product states, say $\hat\rho_A^{out} \otimes\hat\rho_B^{out}$. Let the joint recorded observations of $A$ and $B$ corroborate one such product state in each run of the experiment. This evidently implies that each particular sample of $A$ and $B$ must undergo an evolution from the initial, pre-measurement state $|\Psi_{AB}\rangle$ at $t_0$ to some final, post-measurement state $\hat\rho_A^{out} \otimes\hat\rho_B^{out}$ at $t_0 + \delta t$, occurring with a certain probability among a set of possible product states. System $A$, in particular, must evolve from a local state $\hat\rho_A^{in} = Tr_B \left[ |\Psi_{AB}\rangle\langle \Psi_{AB}| \right ]$ at $t_0$ into the local state $\hat\rho_A^{out}$ at $t_0+\delta t$. But this is exactly the "projection at-a-distance" process that was to be avoided. As usual with the "projection at-a-distance" paradox, one can only notice that for sufficiently small $\delta t$ the recorded observations of $A$ and $B$ become space-like separated events, and in suitably chosen referentials the observation of $A$ can precede the preparatory measurement on $B$. Hence the claim of a causal relationship between the two cannot be given any empirical support, and the interpretation that a local measurement can produce an "instantaneous disentanglement" of an entangled state into a "probabilistic mixture" is rendered meaningless. It is, however, fair to say that a local measurement can produce a "disentanglement" of the total density matrix into a separable form, even if the corresponding mixture cannot  be characterized as "probabilistic". 

It is worth pointing out that the above line of reasoning applies as well if we begin with the assumption that local measurements produce local "probabilistic mixtures". Suppose that the preparatory measurement by a suitable measuring device leaves system $B$ in a "probabilistic mixture" of local states. Then the total state must be a separable mixture of states of $B$ and joint states of $A$ and of the measuring device. But since system $A$ and the measuring device at the location of $B$ were initially in mutually uncorrelated states, and a local operation cannot produce nonlocal entanglement, it follows that the total post-measurement state must be separable with respect to all contributing parties. That is, it must be a statistical superposition of products of local states of $A$, $B$ and the measuring device, respectively. Since this separable form corresponds to a "probabilistic mixture" for $B$, it must necessarily correspond also to a "probabilistic mixture" for both $A$ and the measuring device. Obviously, we have just recovered the remote preparation process under discussion, and the rest of the above argument applies identically. 

It becomes evident now that the statement {\it ' local measurements produce local and/or remote "probabilistic mixtures"' }is equivalent to the "projection postulate". According to this logic, we are presented with the following orthogonal alternatives:\\

(L) We accept the "probabilistic mixture" interpretation of output states of quantum measurements, and thus we implicitly accept the "projection postulate", with all inherent difficulties. On the other hand, the "projection postulate" validates the phenomenon of remote preparation or "projection at-a-distance", which in turn can be invoked to argue that quantum dynamics is necessarily linear.\\

(NL) We circumvent the interpretation of quantum measurement outputs as "probabilistic mixtures", to the effect that mixed quantum states, including said measurement outputs, are to be treated as "elementary mixtures". In this case the concept of "remote preparation" is rendered untenable, and although "elementary mixtures" are {\it statically equivalent} to corresponding "probabilistic mixtures", there no longer is definite support for their {\it dynamical equivalence}. Hence a linear quantum dynamics is no longer indispensable.\\

The first option (L) has been reviewed in the preceding Section. For the sake of a closed argument, we find it instructive to explore here the question whether the second point of view (NL) can somehow yield viable nonlinear dynamical alternatives. Aside from a redefinition of mixed quantum states as {\it elementary mixtures}, we have in mind only an extension of the dynamical law, while all other "static" notions [including observables] are retained in their usual form. 

In this case, we must acknowledge that there already exists considerable evidence, both experimental \cite{tests-1,tests-2,tests-3,tests-4,tests-5,tests-6} and theoretical \cite{Mielnik}, {\it against a nonlinear dynamics of pure states }in closed [isolated] systems. In truth, the past decade has brought us exceptional experimental support \cite{tests-1,tests-2,tests-3,tests-4,tests-5,tests-6} in favor of a linear [and unitary] propagation of pure states. Let us accept this fact as an empirical constraint on all acceptable quantum dynamics, which we cast in the form of a {\it pure state condition: all pure states in the Hilbert space of states of a closed [isolated] quantum system evolve in time according to a linear dynamics.}

The immediate question we face now is: once accepted the pure state condition, {\it does it make any sense whatsoever to consider yet the possibility of a nonlinear dynamics, or is the problem already closed? }

Counter-intuitive as may seem to the contemporary physicist heavily trained in {\it linear dynamics}, the fact that 
pure states must propagate linearly does not preclude, in itself, an intrinsically nonlinear dynamics for mixed states, and so does not imply Eq.(\ref{eq2}) once decided that mixed quantum states must be regarded as "elementary mixtures". In fact, the literature already offers a number of positive, trace preserving nonlinear dynamics that reduce to a simple unitary evolution on {\it all }pure states of the state space \cite{Jordan,Czachor-1,Czachor-2,Czachor-3,Czachor-4,Czachor-5,Czachor-6,Czachor-7,Beretta-1,Beretta-2,EQD}. 

A simple such example is provided by the nonlinear von Neumann equation \cite{Czachor-3} 

\begin{equation}
\label{eq3}
i\hbar\; \dot{\hat\rho} = \left[{ H \left[{ \frac{\hat\rho}{Tr\hat\rho} }\right]^q + \left[{ \frac{\hat\rho}{Tr\hat\rho} }\right]^q H\;,\; \hat\rho }\right]\;\;,
\end{equation}

\noindent with $q>0$ a real scalar, and $H$ a self-adjoint operator. This dynamics produces an evolution of the form

\begin{equation}
\label{eq4}
\hat\rho(t) = S_{\hat\rho}(t) \hat\rho(0) \left[{S_{\hat\rho}(t)}\right]^{\dagger}\;\;,
\end{equation}

\noindent where the propagator $S_{\hat\rho}(t)$ is a unitary operator with a nonlinear dependence on $\hat\rho$, and satisfies the equation of motion

\begin{equation}
\label{eq5}
i\hbar\;\dot{S}_{\hat\rho}(t) =  \left[{ H \left[{ \frac{\hat\rho}{Tr\hat\rho} }\right]^q + \left[{ \frac{\hat\rho}{Tr\hat\rho} }\right]^q H}\right] \cdot S_{\hat\rho}(t)\;\;.
\end{equation}

\noindent One can easily check that pure states evolve into pure states, since $(d/dt)(\hat\rho^2-\hat\rho) \equiv \dot{\hat\rho}\hat\rho + \hat\rho\dot{\hat\rho} - \dot{\hat\rho} = 0$. Moreover, in this case [$\hat\rho^2=\hat\rho$] the nonlinearity disappears, and the von Neumann equation reduces to the familiar linear form

\[
i\hbar\; \dot{\hat\rho} = \left[{ H\;,\; \hat\rho }\right]\;\;.
\]

\noindent The dynamics described by Eq.(\ref{eq4}) becomes physically meaningful if, in addition, it complies with the no-signaling condition. In particular, the uncorrelated states of two noninteracting systems, $A$ and $B$, must propagate into uncorrelated states [separability], and entangled mixed states that produce identical local initial conditions must generate identical local evolutions [locality]. To reconcile these requirements, one may adapt Polchinski's conjecture \cite{Polchinski} [see also \cite{Czachor-1,Czachor-2,Czachor-Doebner}], and postulate the total propagator as the product of local propagators for the local reduced states, i.e., 

\[ 
S_{\hat\rho_{AB}}^{(AB)}(t) = S_{Tr_B \hat\rho_{AB}}^{(A)} (t) \otimes S_{Tr_A \hat\rho_{AB}}^{(B)} (t)\;.
\]
 
\noindent Accordingly, the total generator must read

\begin{equation}
\label{eq6}
G^{(AB)}\left({ \hat\rho_{AB}}\right) =G^{(A)}\left({ Tr_B\hat\rho_{AB}  }\right) + G^{(B)}\left({ Tr_A\hat\rho_{AB} }\right). 
\end{equation} 

\noindent Despite its artificial aspect, this definition finds a natural justification if the density matrix in the original equation of motion (\ref{eq3}) is interpreted from the beginning as a local state obtained by averaging out possible entanglement with a noninteracting environment [$\hat\rho \to \hat\rho_{loc} = Tr_{env}[\hat\rho_{loc+env}]$]. It was argued \cite{Czachor-Doebner} that such well-behaved nonlinear dynamics may indeed be formulated in a manner that avoids both the "wavefunction collapse" paradigm of quantum measurement theory, and a conflict with quantum nonlocality. In a separate paper \cite{nlin-th} we address the problem of an extended framework for the formulation and analysis of these theories .

Here we remark that as a direct corollary of the above observation the problem of linearity in quantum dynamics cannot be considered closed in absence of a test of linear propagation for mixed states. Although this task may seem considerably more demanding than a test of pure state dynamics, we wish to point out that it actually amounts to a test of linearity for the dynamics of entangled pure states. Indeed, the following simple argument shows that no nonlinear theory may account for the linear propagation of both the pure states of isolated systems and the entangled pure states of noninteracting systems without a conflict with the no-signaling condition. Consider again two noninteracting systems $A$ and $B$, and suppose that all pure states of the joint system $(A+B)$ evolve linearly, under a common propagator. By the same reasoning used previously for the case of {\it probabilistic mixtures}, uncorrelated pure states must evolve into uncorrelated pure states and the overall linear propagator must be a direct product of individual linear propagators for $A$ and $B$. If one considers now an arbitrary entangled state, it follows that the corresponding reduced states must also evolve linearly. On the other hand, there are an infinity of entangled {\it mixed }states that produce the same reduced state for either one of the systems. But if entanglement is not to be distilled locally [the no-signaling condition], a given reduced state must evolve in the same manner, regardless of any entanglement with a remote counterpart. Hence the reduced states generated by entangled mixed states must also evolve linearly, which means in effect that the overall dynamics must be linear. 

Conversely, any nonlinear dynamics that is well-behaved with respect to the no-signaling condition, and also conforms to the {\it pure state condition}, must necessarily predict that pure entangled states of noninteracting systems must evolve, and perhaps decohere into mixed states, in a nonlinear manner. We are led in this way to the linearity criterion stated in the opening paragraph, and which we recall here for convenience: \\

{\it \underline{Linearity criterion}: If all pure states of nonseparable, isolated systems evolve linearly, then the overall quantum dynamics is linear if and only if the entangled pure states of noninteracting systems evolve linearly}. \\

This is our basic result concerning the previously defined alternative (NL), under the constraint of the {\it pure state condition}. 

\subsection{Summary}
\label{Sec4}

We showed that, contrary to the assertion of ref.\cite{Gisin}, the relativistic "no-signaling condition" is not necessary for a rationalization of linear quantum dynamics from "quantum statics". It turns out that the "probabilistic mixture" interpretation of mixed quantum states in its dynamical form is already sufficient for this purpose. This is also the essential obstacle faced by nonlinear quantum theories, and not the incompatibility between a nonlinear quantum dynamics and relativity, as advocated lately. Perhaps it may be argued that the Hilbert space structure of the quantum state space is determined by the relativistic structure of space-time \cite{Svetlichny}. But once this structure is accepted as the fundamental component of {\it quantum statics}, and is supplemented by the "probabilistic mixture" point of view, there is no need for a recurring reference to relativity in order to justify the linearity of quantum dynamics. 

On the other hand, the "probabilistic mixture" interpretation of mixed states implies a covert acceptance of the "projection postulate". If one chooses to dismiss "projections-at-a-distance", e.g., as incompatible with relativity, then mixed quantum states must be regarded exclusively as "elementary mixtures". A suitable physical argument for this point of view may be, for instance, that there always exists a finite chance of nonvanishing, nondistillable entanglement with noninteracting components of the environment. In this case previous arguments \cite{Gisin,nlin-ftl-1,nlin-ftl-2,nlin-ftl-3,Polchinski} that rule out dynamical nonlinearity by conflict with the dynamical "probabilistic mixture" interpretation under the "no-signaling condition" loose their ground, and well-behaved, relativistically consistent nonlinear dynamics become possible \cite{Jordan,Czachor-1,Czachor-2,Czachor-3,Czachor-4,Czachor-5,Czachor-6,Czachor-7}.

Certainly, the viability of these theoretical alternatives is essentially limited by empirical evidence. We have given particular attention to the very likely empirical constraint \cite{tests-1,tests-2,tests-3,tests-4,tests-5,tests-6} that all pure states of isolated systems must evolve unitarily into pure states, as in linear quantum theory. An important result concerning this possibility is a simple {\it linearity criterion}, which suggests that a decisive empirical answer to the (non)linearity problem in quantum dynamics can be provided in two steps: \\

i) a test of linear [and unitary] propagation of pure states in closed systems, which is already available \cite{tests-1,tests-2,tests-3,tests-4,tests-5,tests-6};\\

ii) a test of linear [and unitary] propagation of entangled pure states of noninteracting systems.


\begin{thebibliography}{00}


\bibitem{Gisin} C. Simon, V. Buzek, N. Gisin, Phys. Rev. Lett. {\bf 87} (2001) 170405 .

\bibitem{Bona} P.Bona, preprint quant-ph/0201002.

\bibitem{Bona-1} P.Bona, Acta Phys. Slov. {\bf 50} (2000) 1; also preprint math-ph/990022, quant-ph/9910011.

\bibitem{Gyftopoulos} G.N. Hatsopoulos and E.P. Gyftopoulos, Found.Phys. {\bf 6} (1976) 15, 561.

\bibitem{nlin-ftl-1} N. Gisin, Phys.Lett. A {\bf 143} (1990) 1.

\bibitem{nlin-ftl-2} M. Czachor, Found.Phys.Lett. {\bf 4} (1991) 351.

\bibitem{nlin-ftl-3} T.F. Jordan and Z.-E. Sariyanni, Phys. Lett. A {\bf 263} (1999) 263.

\bibitem{Polchinski} J. Polchinski, Phys.Rev.Lett {\bf 66} (1991) 397.

\bibitem{Mielnik} B. Mielnik, preprint quant-ph/0012041; Phys.Lett. A {\bf 289} (2001) 1.

%\bibitem{disentgl} D. Terno, Phys.Rev. A {\bf 59} (1999) 3320. T. Mor, Phys. Rev. Lett. {\bf 83}, 1451 (1999). T. %Mor and D. Terno, Phys.Rev. A{\bf 60} (1999) 4341. 

\bibitem{Jordan} T.F. Jordan, Ann. Phys. (N.Y.) {\bf 225} (1993) 83. 

\bibitem{Czachor-1} M. Czachor, Phys.Rev. A {\bf 57} (1998) 4122. 

\bibitem{Czachor-2} M. Czachor and M.Kuna, Phys.Rev. A {\bf 58} (1998) 128. 

\bibitem{Czachor-3} S.B. Leble and M. Czachor, Phys.Rev. E {\bf 58} (1998) 7091. 

\bibitem{Czachor-4} M.Czachor and J. Naudts, Phys.Rev. E {\bf 59} (1999) 2497. 

\bibitem{Czachor-5} M. Czachor, Int.J.Theor.Phys. 38 (1999) 475-500. 

\bibitem{Czachor-6}M. Czachor, M. Kuna, S. B. Leble, J. Naudts, "Nonlinear von Neumann-type equations", in "New insights in quantum mechanics", edited by H.D. Doebner, S.T. Ali, M. Keyl, and R.F. Werner (World Scientific, 1999).

\bibitem{Czachor-7} N. Ustinov, M. Czachor, M. Kuna, S.B. Leble, Phys. Lett. A {\bf 279} (2001) 333. 

\bibitem{Beretta-1} G.P. Beretta, E.P. Gyftopoulos, J.L. Park, and G.N. Hatsopoulos, Nuovo Cimento Soc. Ital. Fis., B {\bf 82} (1984) 169.

\bibitem{Beretta-2} G.P. Beretta, Found.Phys. {\bf 17} (1987) 365. 

\bibitem{EQD} S. Gheorghiu-Svirschevski, Phys.Rev. A {\bf 63} (2001) 022105.

\bibitem{tests-1} J.J.Bollinger et al., Phys.Rev.Lett.\textbf{63} (1989) 1031.

\bibitem{tests-2} R.L. Walsworth, F. Silvera, E.M. Mattison and R.F.C. Vessot, Phys.Rev.Lett. \textbf{64} (1990) 2599.

\bibitem{tests-3} T.E. Chupp and R.J. Hoare, Phys.Rev.Lett. \textbf{64} (1990) 2261.

\bibitem{tests-4} P.K. Majumder et al., Phys.Rev.Lett. \textbf{65} (1990) 2931.

\bibitem{tests-5} F.Benatti and R.Floreanini, Phys.Lett.\textbf{B389} (1996) 100.

\bibitem{tests-6} F.Benatti and R.Floreanini, Phys.Lett.\textbf{B451} (1999) 422.

\bibitem{Czachor-Doebner} M. Czachor, H.D. Doebner, preprint quant-ph/0110008; quant-ph/0106051.

\bibitem{nlin-th} S. Gheorghiu-Svirschevski, quant-ph/0207***. 

\bibitem{Svetlichny} G. Svetlichny, Found.Phys. {\bf 28}, (1998) 131; {\bf 30} (2000) 1819; also preprint quant-ph/9512004. 

\end{thebibliography}
\end{document}